# In Vivo Quantification of Arterial Active Mechanics Using Deep Learning-Assisted Pressure-Area Analysis

Yuxuan Jiang [1], Yanping Cao [1, *]

[1] Institute of Biomechanics and Medical Engineering, AML, Department of Engineering Mechanics, Tsinghua University, Beijing 100084, China

[*] Correspondence: Yanping Cao (caoyanping@tsinghua.edu.cn)

## ABSTRACT

Active arterial mechanics, governed by vascular smooth muscle contraction, are critical to physiological regulation, cardiovascular disease progression, and clinical diagnosis. Although various in vivo methods have been developed to assess arterial stiffness, most cannot distinguish the contribution of smooth muscle tone; therefore, quantitative characterization of arterial activity remains challenging. In this study, we developed a pressure–area analysis framework integrating ultrasound imaging, blood pressure measurement, neural network–based segmentation of arterial cross-sectional area, and biomechanical model–driven inversion to infer active mechanical properties. A total of 233 volunteers (aged 18 – 65 yr.) were recruited to acquire cross-sectional ultrasound videos of the right common carotid artery for training the neural network. The segmentation results demonstrate good spatial and temporal performance of the neural network. We further recruited 10 additional volunteers (aged 25 ± 3 yr.) to perform a 1-minute step test, followed by pressure-area measurements over a 30-minute recovery period. Using the proposed approach, we quantified post-exercise changes in carotid arterial active mechanics relative to baseline (i.e., the resting state). Results showed that active mechanics remained elevated for approximately 15 minutes compared to baseline ($p < 0.05$), whereas systolic pressure differed significantly only within the first approximately 5 minutes post-exercise ($p < 0.001$). These results indicate a dissociation between blood pressure and smooth muscle recovery, which may offer new insight into vascular smooth muscle regulation during physiological stress.

# 1 Introduction

Arterial stiffness plays a pivotal role in cardiovascular health and disease. It is a strong predictor of adverse clinical outcomes, such as hypertension, stroke, and heart failure (Boutouyrie et al. 2021; Cecelja and Chowienczyk 2012; Chirinos et al. 2019). In conventional clinical assessments (Laurent et al. 2006; Segers et al. 2020; Teixeira et al. 2016), arterial stiffness is typically regarded as a passive mechanical property determined largely by the extracellular matrix (ECM), which comprises structural proteins such as elastin and collagen (Holzapfel et al. 2000). However, arteries are active biological tissues, with vascular smooth muscle cells (SMCs) capable of dynamically regulating vascular stiffness (Humphrey and Na 2002; Rachev and Hayashi 1999). Unlike ECM, which contributes to the long-term structural integrity, SMCs can rapidly alter vessel tone in response to neural, humoral, and mechanical stimuli (Touyz et al. 2018). Emerging evidence suggests that altered smooth muscle tone may contribute to early vascular dysfunction in various conditions, including essential hypertension (Rizzoni and Rosei 2006), diabetes (Schofield et al. 2002), and aging (Harvey et al. 2015). Reduced SMC responsiveness has been observed in diseases such as autonomic neuropathy (D Lefrandt et al. 2010) and heart failure with preserved ejection fraction (Alpenglow et al. 2024). These findings highlight the potential of arterial active mechanics as a functional biomarker, and underscore the need for accurate assessment of this dynamic property in vivo.

Beyond its clinical implications, arterial activity also plays a fundamental role in physiological regulation. Active mechanical changes in smooth muscle cells are involved during and after exercise, stress, and other autonomic stimuli, contributing to short-term variation of arterial stiffness (Reesink and Spronck 2019). For instance, resistance or isometric exercise often leads to temporary increases in stiffness, with recovery typically occurring within minutes to hours (DeVan et al. 2005; Jurik et al. 2021; Studinger et al. 2003; Zhang et al. 2021). These short-term fluctuations reflect vascular reactivity, autonomic regulation and functional adaptation, positioning smooth muscle behavior as a key focus in both physiological studies and biomedical research. As such, reliable methods for quantifying arterial active mechanics in vivo are essential not only in clinical diagnostics but also in basic physiology and translational studies.

Although much effort has been made over past years, accurate *in vivo* characterization of arterial activity remains challenging. Current approaches for assessing arterial stiffness, including pressure-area analysis (Chirinos 2012; Jiang et al. 2022; Khamdaeng et al. 2012; Spronck et al. 2015; Valdez-Jasso et al. 2011), guided wave elastography (Couade et al. 2010; Jiang et al. 2025; Li et al. 2022), and pulse wave velocity measurement (Salvi et al. 2019; Van Bortel et al. 2012), primarily evaluate integrated arterial wall stiffness without effectively distinguishing contributions from active smooth muscle contraction. Recent studies have attempted to infer arterial active mechanics using pressure-area curves (Gade et al. 2021; Masson et al. 2008; Pewowaruk and Gepner 2022; Reesink and Spronck 2019); however, limited experimental data, coupled with constitutive modeling complexity and parameter fitting challenges, present significant obstacles to precise quantification of smooth muscle contributions. Additionally, precise arterial area measurement poses further difficulties. While ultrasound M-mode imaging estimates vessel diameter (Masson et al. 2008; Valdez-Jasso et al. 2011), its accuracy is critically dependent on probe alignment—off-center A-lines may substantially underestimate diameter. Image segmentation algorithms have been proposed for cross-sectional arterial area estimation (Azzopardi et al. 2020; Golemati et al. 2007; Guerrero et al. 2007; Noble and Boukerroui 2006; Wang et al. 2009); however, achieving temporally smooth and accurate area waveforms across cardiac cycles remains challenging.

To overcome above challenges, this study proposes an integrated framework to infer arterial active mechanics by combining noninvasive ultrasound imaging, blood pressure measurements, deep learning-based arterial segmentation, and biomechanical model-driven inversion. Relying on this framework, we establish a non-invasive method for measuring the dynamic variations of active mechanical parameters in arteries. *In vivo* post-exercise experiments validated the method's capacity to track transient arterial activity changes over approximately 30 minutes following a 1-minute step test. The potential applications of this method in vascular physiology and clinical diagnosis are discussed.

## 2 Materials and methods

### 2.1 Dataset for deep learning-based vessel segmentation

We recruited 233 volunteers (aged 18 - 65 years old, 155 males and 78 females) and acquired transverse B-mode ultrasound videos of their right common carotid arteries. All recorded arteries were free of atherosclerotic plaques. The ultrasound examinations were performed using a clinical ultrasound imaging system (Clover 60, Wisonic, China) equipped with a linear array transducer (L15-4, center frequency 9.5 MHz). Prior to the examination, each participant signed an informed consent. Subjects were instructed to rest calmly for at least 5 minutes before the measurement and were examined in a seated position. During scanning, the ultrasound probe was gently placed on the skin surface to avoid excessive compression of the artery. Continuous B-mode ultrasound imaging was recorded for approximately 5 seconds at a frame rate of 83 frames per second. After the ultrasound measurement, each subject's brachial blood pressures were measured using an electronic blood pressure monitor (HEM-7200, Omron, Japan). The demographic information of the participants is listed in Table 1. The protocol was approved by the institutional review board at Tsinghua University (Project No: THU01-20250057).

**Table 1.** Demographic information of participants recruited in the two in vivo experiments

|  | Dataset for vessel segmentation | Post-exercise experiment |
|---|---|---|
| Number of participants | 233 (155 males, 78 females) | 10 (10 males) |
| Age (years old) | 30 ± 13 | 25 ± 3 |
| BMI (kg/m$^2$) | 22.4 ± 3.0 | 22.2 ± 1.8 |
| Brachial diastolic blood pressure (mmHg) | 71 ± 9 | 70 ± 9 (baseline) |
| Brachial systolic blood pressure (mmHg) | 116 ± 14 | 113 ± 7 (baseline) |

## 2.2 Convolutional neural network for vessel segmentation

A total of 233 ultrasound videos were used for training and evaluation of the neural network. These videos were randomly partitioned into training (n = 206), validation (n = 14), and test (n = 13) sets. All the videos underwent preprocessing to remove overlaid textual information and were cropped to retain only the region of interest around the arterial lumen, resulting in square images of 448 × 448 pixels. Temporally, each video was further trimmed to a duration of 3.5 seconds. Under the supervision of an experienced sonographer, the arterial lumen was annotated to serve as ground truth for segmentation model training. A weak supervision strategy was adopted, i.e., for each video, only two frames were manually labeled— including end-diastole (minimum) and peak-systole (maximum) lumen areas. To simplify the annotation process, the lumen boundary was approximated using a circular shape. Despite the use of simplified circular annotations, the results demonstrated the capability of the model to predict anatomically variable lumen shapes.

Inspired by previous work on image segmentation of the left ventricle (Ouyang et al. 2020), we adopted the DeepLabv3 architecture (Chen et al. 2017) for vessel segmentation. This architecture is based on convolutional neural networks (CNN) and incorporates atrous convolutions to capture multi-scale contextual information, as well as residual connections to facilitate gradient flow and improve feature propagation in deeper layers (Fig. 1a). The training objective was to minimize the pixel-wise binary cross-entropy loss. The model was initialized with random weights and trained using a stochastic gradient descent optimizer. Training was conducted for 50 epochs with an initial learning rate of $10^{-6}$, a momentum of 0.9, and a batch size of 4. The model was implemented in Python using PyTorch and trained on a single NVIDIA GeForce RTX 2080 Ti GPU. During training, the model input was the two annotated frames from each video. Once trained, the model was applied to the full sequence of the interested video to predict the temporal variation of the arterial lumen area on a frame-by-frame basis.

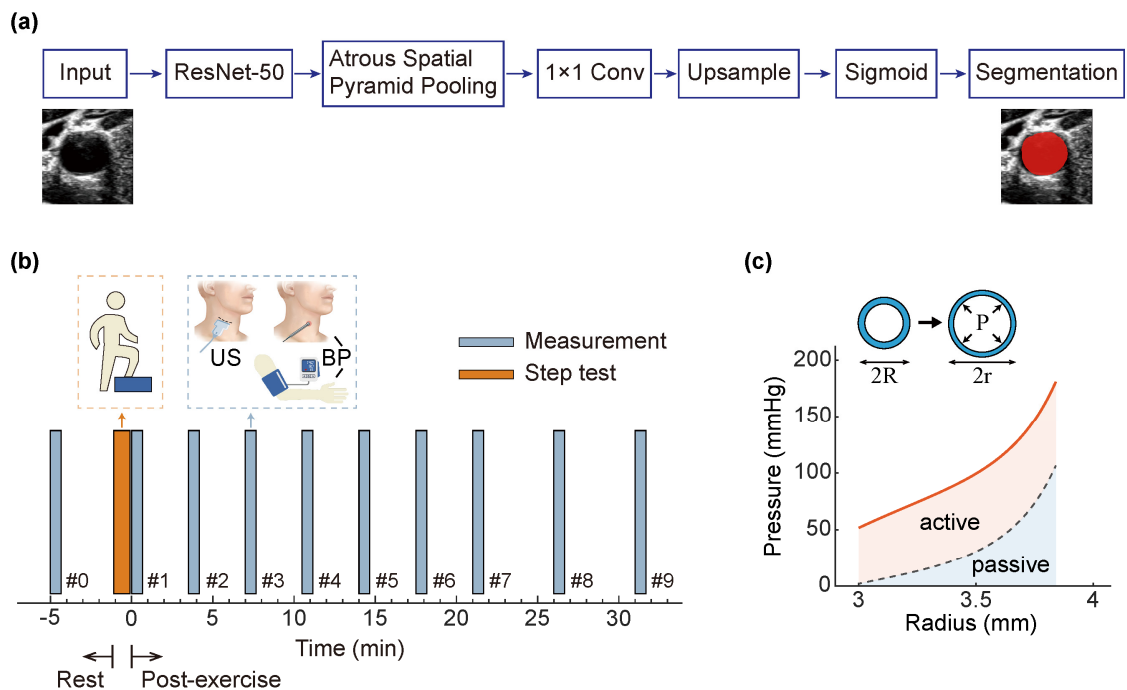

**Figure 1.** Vessel segmentation framework and in vivo experimental setup for post-exercise assessment. (**a**) Convolutional neural network framework for vessel segmentation. (**b**) Experimental procedure for in vivo measurements of blood pressure and arterial ultrasound pre- and post-exercise. US, Ultrasound; BP, Blood pressure measurements. (**c**) The theoretical pressure–radius relationship of arteries (via Eq. (5)), including both active and passive components.

## 2.3 In vivo post-exercise experiment

Ten healthy young male volunteers (23±3 years old) were recruited for the post-exercise study. They were excluded from the vessel segmentation dataset. The demographic information of the participants is listed in Table 1. All subjects signed an informed consent form before the experiments. For each volunteer, ultrasound and blood pressure measurements of the right common carotid artery were first performed under resting conditions (denoted as #0). The volunteer then performed a 1-minute step test, stepping up and down on a 30 cm-high platform at a frequency of one complete cycle every 2 seconds. Immediately after the exercise (at time 0, denoted as #1), ultrasound and blood pressure measurements were conducted again. Subsequently, follow-up measurements were performed every 3 minutes and 30 seconds for six time points (#2 - #7), followed by another two measurements at 5-minute intervals (#8 - #9). In total, nine post-exercise measurements were conducted over approximately 30 minutes. Including the baseline measurement at rest, each session consisted of 10 measurement time points. Figure 1b illustrates the experimental procedure described above. During the entire pre- and post-exercise monitoring period, participants remained in a seated position. The head was kept in a neutral position during measurements to avoid neck extension. The left arm was placed on a table at heart level to ensure accurate brachial blood pressure measurement.

For the measurement at each time point, cross-sectional B-mode ultrasound images of the right common carotid artery were acquired using an ultrasound imaging system (Clover 60, Wisonic, China) equipped with a linear array transducer (L15-4, central frequency: 9.5 MHz). The acquisition lasted approximately 5 seconds at a frame rate of 83 frames per second. Care was taken to avoid excessive compression of the artery by the ultrasound probe during scanning. Immediately after the ultrasound acquisition, carotid arterial pressure at the same anatomical location was recorded using an applanation tonometer (SPT-301, Millar Inc., USA) at a sampling rate of 1 kHz for approximately 5 seconds. Concurrently, brachial systolic and diastolic blood pressures were measured on the left arm using a cuff-based electronic blood pressure monitor (HEM-7200, Omron, Japan). The brachial blood pressure values were used to calibrate the pressure waveform acquired via applanation tonometry (details are provided in Section 2.4). The total duration of each measurement was approximately 40 seconds, including ultrasound imaging, applanation tonometry, and cuff-based blood pressure acquisition.

Each subject underwent this protocol three times on three separate days. The protocol was approved by the institutional review board at Tsinghua University (Project No: THU01-20250057).

## 2.4 Calibration of arterial blood pressure

The carotid pressure waveform obtained by applanation tonometry was calibrated by the Kelly and Fitchet method (Kelly and Fitchett 1992; Laurent et al. 2006; Mitchell et al. 2010). This calibration is based on the principle that the diastolic and mean blood pressure is constant through the large artery tree. Brachial diastolic (DBP) and systolic blood pressure (SBP) were measured using the electronic blood pressure monitor. Brachial mean blood pressure (MBP) were calculated using MBP = SBP/3 + 2×DBP/3 (Zakrzewski and Anthony 2017). The mean blood pressure at the common carotid artery is calculated by $P_{\text{avg}} = \int_0^T P(t)\mathrm{d}t\,/\mathrm{T}$. With the use of relationship $P_{\text{avg}} = \text{MBP}$ and $P_{\min} = \text{DBP}$, the blood pressure at carotid artery, $P(t)$, can be determined. More details can be found in our previous work (Jiang et al. 2025; Jiang et al. 2022).

## 2.5 Measurement of arterial lumen area

The ultrasound B-mode videos acquired from the post-exercise experiments were input into the trained neural network for frame-by-frame segmentation of the arterial lumen. This yielded continuous area data over time, denoted as $A(t)$. The arterial wall thickness at end-diastole (denoted as $h$) was also manually measured from the B-mode images.

To obtain the arterial pressure–area curve, the pressure data was down-sampled to match the temporal resolution of the area measurement (i.e. 83 Hz). Since the pressure and ultrasound measurements were not acquired synchronously, temporal alignment was required. Based on previous literature (Jiang et al. 2022; Masson et al. 2008), it is assumed that the pressure and area curves are in phase; therefore, the two signals were aligned by matching their respective peaks.

## 2.6 Mechanical modeling

The artery is modeled as a thin-walled elastic cylinder. Two configurations are defined: the initial (stress-free) configuration and the deformed (loaded) configuration, where the artery is subjected to the internal blood pressure $P$ and longitudinal stretch. In the initial configuration, the artery has a radius $R$, wall thickness $H$, and longitudinal length $L$; in the deformed configuration, these become $r$, $h$, and $l$, respectively. With the incompressible assumption of the artery material, this geometric constraint holds: $RHL = rhl$. A cylindrical coordinate system $(\bar{r}, \theta, z)$ was built on the model, where $\bar{r}$ denotes the radial direction, $\theta$ the circumferential direction, and $z$ the longitudinal direction. The stretch ratios in the radial, circumferential, and longitudinal directions are denoted as $\lambda_{\bar{r}}$ ($= h/H$), $\lambda_\theta$ ($= r/R$), and $\lambda_z$ ($= l/L$), respectively. The incompressible condition yields: $\lambda_{\bar{r}} \lambda_\theta \lambda_z = 1$.

Under the thin-walled assumption, the relationship between the blood pressure and deformation is given by (Ogden 2001)

$$P = \frac{h}{r}\sigma_\theta = \frac{h}{r}\lambda_\theta \frac{\partial W}{\partial \lambda_\theta} \tag{1}$$

where $\sigma_\theta$ denotes the circumferential Cauchy stress. $W$ is the strain energy function of the artery. The strain energy function is expressed as the sum of a passive component ($W^{(p)}$) and an active component ($W^{(a)}$), i.e.,

$$W = W^{(p)} + W^{(a)} \tag{2}$$

The passive part characterizes the nonlinear mechanical response arising from structural constituents including collagen fibers, elastin fibers, and extracellular matrix. The active part accounts for the contractile behavior of vascular smooth muscle cells.

Various passive arterial constitutive models have been proposed, such as two-fiber (Gasser et al. 2006; Holzapfel et al. 2000; Holzapfel et al. 2015; Holzapfel and Ogden 2010) and four-fiber family models (Baek et al. 2007). In this study, a two-fiber family model is adopted, with the fibers assumed to be aligned in the longitudinal and circumferential directions, respectively. The strain energy function is

$$W^{(p)} = \frac{\mu}{2}(I_1 - 3) + \frac{\mu_\theta}{2k_2}\left\{\exp\left[k_2(\lambda_\theta^2 - 1)^2\right] - 1\right\} + \frac{\mu_z}{2k_2}\left\{\exp\left[k_2(\lambda_z^2 - 1)^2\right] - 1\right\} \quad (3)$$

where the first term on the right-hand side describes the isotropic elastin matrix. The second and third terms represent two families of collagen fibers oriented in the circumferential and longitudinal directions, respectively. $\mu$ denotes the shear modulus of the matrix. $\mu_\theta$ and $\mu_z$ represent the collagen fiber-related shear moduli along the circumferential and longitudinal directions, respectively. $k_2$ is a dimensionless parameter denoting the nonlinear hardening effect of the collagen fibers. The first principle invariant $I_1 = \lambda_\theta^2 + \lambda_z^2 + \lambda_{\bar{r}}^2$.

A number of arterial active constitutive models have been proposed in the literature to describe the relationship between the active force and the contractile deformation of vascular smooth muscle cells, such as the phenomenological models (Baek et al. 2007; Franchini et al. 2022; Wagner and Humphrey 2011; Zulliger et al. 2004) and the mechanochemical models (Murtada et al. 2010a; Murtada et al. 2010b; Schmitz and Böl 2011; Stålhand et al. 2008). In this study, we adopt a relatively simple active model with clear mechanical interpretation (Baek et al. 2007; Humphrey and Na 2002); the strain energy function is

$$W^{(a)} = K_a\left[\lambda_f + \frac{1}{3}\frac{(\lambda_m - \lambda_f)^3}{(\lambda_m - \lambda_0)^2}\right] \quad (4)$$

where $K_a$ represents an activation function with units of stress, intrinsically linked to intracellular calcium concentration — higher calcium levels lead to stronger activation. $\lambda_m$ denotes the stretch at which active force generation reaches its maximum, while $\lambda_0$ denotes the minimum stretch below which no active contraction occurs. $\lambda_f$ is the stretch of the smooth muscle fibers. Following the approximation adopted in previous studies (Humphrey and Na 2002; Masson et al. 2008), the distribution of vascular smooth muscle cells is assumed to align predominantly along the circumferential direction of the artery, leading to the relation $\lambda_f = \lambda_\theta$, which will be used in the subsequent analysis.

Inserting Eqs. (2) - (4) into Eq. (1), we obtain an explicit relation between pressure, deformation and material parameters as follows:

$$\frac{Pr}{h} = \mu(\lambda_\theta{}^2 - \lambda_\theta{}^{-2}\lambda_z{}^{-2}) + 2\mu_\theta \lambda_\theta{}^2(\lambda_\theta{}^2 - 1)\exp\left[k_2(\lambda_\theta{}^2 - 1)^2\right] + K_a\lambda_\theta\left[1 - \left(\frac{\lambda_m - \lambda_\theta}{\lambda_m - \lambda_0}\right)^2\right] \quad (5)$$

where the first two terms on the right-hand side correspond to the passive stress contributions ($\sigma_\theta^{(p)}$) from the stretch of matrix, elastin and collagen fibers, while the third term represents the active stress ($\sigma_\theta^{(a)}$) generated by smooth muscle contraction. Figure 1c shows the theoretically predicted pressure–radius curve. Changes in either the active or passive component can alter the pressure–radius relationship.

## 2.7 Inverse method

According to Eq. (5), eight parameters were involved in the optimization process, including the passive constitutive parameters ($\mu$, $\mu_\theta$, $k_2$), the active parameters ($K_a$, $\lambda_m$, $\lambda_0$), the arterial radius in the stress-free state $R$, and the longitudinal deformation $\lambda_z$. For each individual on each test day, a total of ten pressure–area curves were obtained from the in vivo experiments, denoted as $(P_i^{(I,\exp)}, A_i^{(I,\exp)})$, where the superscript $I$ (= 0, 1, …, 9) indicates the measurement index (corresponding to #0 - #9, respectively). The subscript $i$ (= 1, 2, …, 83) denotes the measured data points at each measurement index.

Joint inversion was performed using all ten pressure-area curves to enhance the robustness and confidence of the estimated parameters. The passive parameters $\mu$, $\mu_\theta$, $k_2$, the active parameters $\lambda_m$, $\lambda_0$, and the initial radius $R$ were assumed to remain constant during the whole measurement period. Based on literature studies (Gade et al. 2021; Jiang et al. 2022; Masson et al. 2011; Masson et al. 2008), the longitudinal stretch ratio $\lambda_z$ also remains constant during the cardiac cycle. Given the above considerations, the only varying parameter across the ten curves was the active parameter $K_a$, denoted as $K_a^{(I)}$ ($I = 0, 1, …, 9$). The optimization function is defined in a form of root-mean-square-error:

$$\mathcal{L} = \frac{1}{10}\sum_{I=1}^{10}\sqrt{\frac{1}{n}\sum_{i=1}^{n}\left(P_i^{(I,\text{theo})} - P_i^{(I,\exp)}\right)^2} \quad (6)$$

where $P_i^{(I,\text{theo})}$ denotes the theoretically predicted pressure, calculated by substituting experimentally measured radius $r_i^{(I,\exp)}$ ($= \sqrt{A_i^{(I,\exp)}/\pi}$), wall thickness $h^{(I,\exp)}$, as well as the unknown parameters $\mu$, $\mu_\theta$, $k_2$, $\lambda_m$, $\lambda_0$, $K_a^{(I)}$, $R$, $\lambda_z$ into Eq. (5). The optimization process was achieved using a particle swarm optimization algorithm (Wang et al. 2018b). The optimal parameter values were identified by minimizing the objective function i.e., $\min(\mathcal{L})$, with a convergence criterion typically satisfied when $\mathcal{L} < 0.1$ in practice. Based on literature data (Baek et al. 2007; Humphrey and Na 2002; Jiang et al. 2022; Masson et al. 2011; Masson et al. 2008), the parameter space is set as $1 < \mu < 500$ kPa, $1 < \mu_\theta < 500$ kPa, $0.1 < k_2 < 20$, $1.6 < \lambda_m < 1.8$, $0.8 < \lambda_0 < 0.9$, $0 < K_a < 300$ kPa, $1 < \lambda_z < 1.3$, $0.7 r_{\min}^{(0)} < R < 0.9 r_{\min}^{(0)}$ (where $r_{\min}^{(0)}$ denotes the minimum radius measured at rest).

After obtaining $K_a$ through curve fitting, the change in the active parameter at each post-exercise time point compared to the baseline is subsequently calculated, i.e. $\Delta K_a^{(I)} = K_a^{(I)} - K_a^{(0)}$ ($I = 1, \ldots, 9$). The parameter $\Delta K_a^{(I)}$ represents the exercise-induced change of smooth muscle activation compared to the baseline. As will be detailed in Section 3.3, this parameter shows greater reliability in parameter inversion compared to $K_a$.

## 2.8 Statistical analysis

Both statistical analysis and parameter estimation process were preformed using MATLAB R2016b (MathWorks Inc., USA). All *in vivo* experimental data were expressed as mean ± standard deviation (s.d.). The paired *t*-test was performed to compare blood pressure and arterial area before and after exercise. A one sample *t*-test was conducted on the parameter $\Delta K_a$ to assess the change in smooth muscle activation before and after exercise. A *p*-value of 0.05 was adopted to indicate statistical significance. Results with $0.01 \leq p \leq 0.05$ were marked with one asterisk (*); those with $0.001 \leq p < 0.01$ were marked with two asterisks (**); and those with $p < 0.001$ were marked with three asterisks (***).

# 3 Results

## 3.1 Arterial lumen segmentation using convolutional neural network

Figure 2a illustrates the segmentation results of the carotid artery lumen using the convolutional neural network (CNN). Each video in the dataset consists of around 300 frames, with manual annotations provided for only two frames (middle row in Fig. 2a). Despite such sparse supervision, the network is able to segment all frames with satisfactory accuracy (bottom row in Fig. 2a). On the test dataset, the CNN achieved a Dice similarity coefficient of 0.96 for the overall segmentations. In comparison, the 4-frame supervision strategy does not show improved segmentation performance (see details in Supplementary Note 1). Consequently, the current approach demonstrates advantages in terms of reduced annotation costs and enhanced training efficiency.

Figure 2b shows the temporal variation of the lumen area for a representative test sample. The black curve represents the result from the CNN-based segmentation, while light-colored curve shows the result of a conventional edge-based method for comparison (see details in Supplementary Note 2). Compared to the edge-based approach, the CNN output exhibits superior temporal smoothness, which enables the detection of fine pulsatile features such as the dicrotic notch, and is particularly critical for accurate reconstruction of pressure–radius curves (see details in Supplementary Fig. S2).

Figures 2c – e further demonstrate the robustness of the CNN method under several challenging conditions. In Fig. 2c, the arterial area exhibits minimal variation over the cardiac cycle (relative change ~8%, below the average level of ~30%). Nevertheless, the CNN still produces smooth and pulsatile area curves. In Fig. 2d, where the ultrasound image exhibits low signal-to-noise ratio (SNR), the edge-based method provides noisy and unstable predictions, while the CNN maintains reliable segmentation (also see Supplementary Video S2). In Fig. 2e, the artery undergoes substantial global motion across frames, primarily due to probe-induced motion. The CNN-based method maintains smooth segmentation results and accurately tracks vessel motion, whereas the edge-based approach leads to noisy predictions (also see Supplementary Video S3). Given that low-SNR images, typically due to insufficient ultrasound coupling gel or subject-specific acoustic attenuation, and motion artefacts from probe movements, subject swallowing, or other body motions are frequently encountered in clinical practice, the CNN method shows strong potential for robust vessel segmentation under clinical imaging conditions.

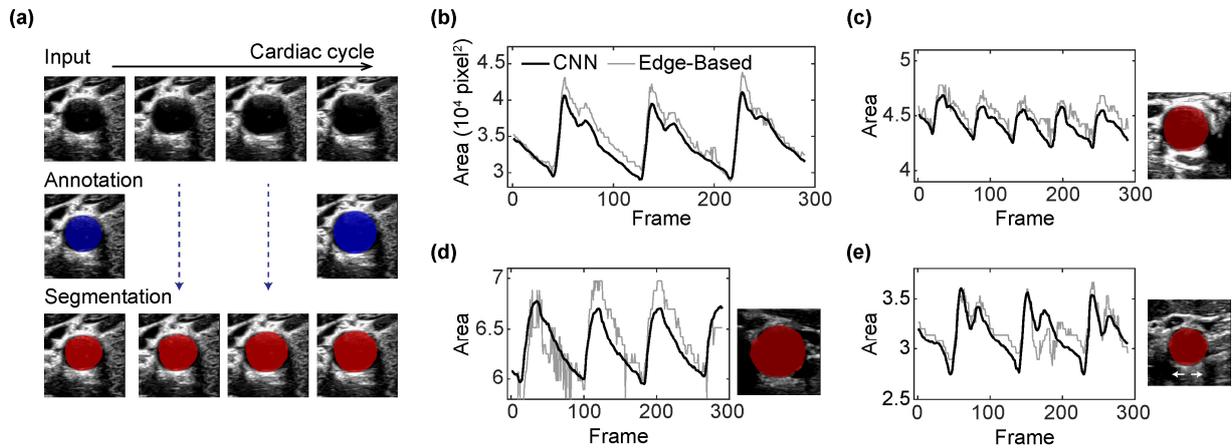

**Figure 2.** Arterial lumen segmentation using convolutional neural network (CNN). (**a**) Segmentation performance of human common carotid arteries on the test dataset. Despite annotations being provided for only two frames in a continuous video, the neural network demonstrates consistently accurate segmentation across all frames. (**b**) Comparison of continuous vessel segmentation performance between the CNN method and a traditional edge-based approach. The analyzed artery corresponds to the one demonstrated in (a). (**c**) – (**e**) Comparison of continuous vessel segmentation performance between the two methods under several challenging conditions, including (c) minimal arterial area variation during the cardiac cycle, (d) low signal-to-noise ratio in the ultrasound images, and (e) obvious vessel motion in the image due to handheld probe movement during imaging. Ultrasound images in (c)–(e) show representative segmentation results using the CNN method. All results presented here are from the test dataset.

### 3.2 Dynamic changes in arterial blood pressure and area after exercise

Figure 3a illustrates the continuous waveforms of carotid artery area and blood pressure in a volunteer (27 years old, male) before and after exercise during a period of ~ 35 minutes. Notably, the area waveform is obtained using the trained neural network, and its smooth temporal profile further confirms the robustness of the network when applied to previously unseen data. As illustrated in Fig. 3a, immediately after exercise (#1), both heart rate and systolic pressure increased markedly compared to the baseline. Over the subsequent 30 minutes, heart rate, blood pressure, and arterial area gradually returned to baseline levels. For each measurement (~3.5 s duration), the average systolic (denoted as $P_{\max}$) and diastolic pressures ($P_{\min}$), as well as the average systolic ($A_{\max}$) and diastolic areas ($A_{\min}$) across multiple cardiac cycles, were calculated. Figure 3b shows the temporal evolution of $P_{\max}$, $P_{\min}$, $A_{\max}$, and $A_{\min}$ before and after exercise. As illustrated, $P_{\max}$ and $A_{\max}$ exhibited substantial variation before and after exercise (relative variation ~ 40% and ~ 12%, respectively), while $P_{\min}$ and $A_{\min}$ showed minimal changes (relative variation < 6%).

Figure 3c presents the statistical results of $P_{\max}$ and $P_{\min}$ across 10 volunteers. A paired *t*-test was performed between each post-exercise measurement and the baseline to assess changes induced by exercise. The results show that systolic pressure remained significantly elevated within ~ 4 minutes post-exercise ($p$ <0.001), while diastolic pressure remained significantly elevated within ~ 8 minutes ($p$ <0.01). Figure 3d shows the statistical results of $A_{\max}$ and $A_{\min}$. Compared to the baseline, $A_{\max}$ significantly increased immediately after exercise ($p$ < 0.001), but showed a significant decrease at 10 – 15 minutes post-exercise ($p$ < 0.01). Notably, blood pressure and arterial area did not follow a simple positive correlation, suggesting the involvement of active regulation by arterial smooth muscle, which will be investigated in the following section.

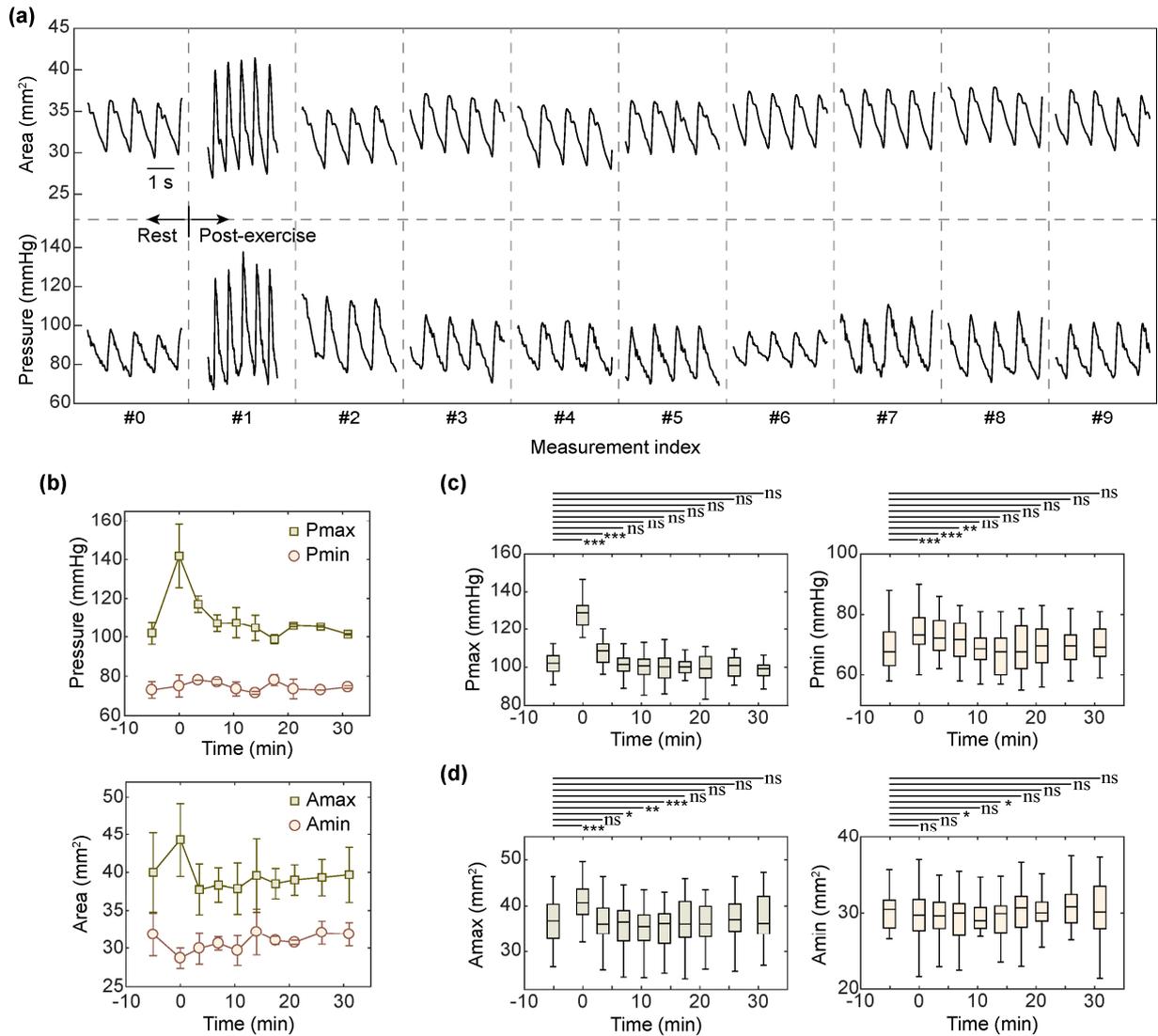

**Figure 3.** Changes in arterial area and blood pressure before and after exercise. (**a**) Temporal changes in arterial area and blood pressure in a representative volunteer. A total of 10 measurements were recorded, including one pre-exercise baseline (#0) and nine post-exercise follow-ups (#1 – #9). Each single measurement lasted ~ 3.5 s. (**b**) Temporal changes in the maximum and minimum values of blood pressure (top) and arterial area (bottom) for the representative volunteer. The maximum and minimum values represent the average systolic and diastolic values computed over multiple cardiac cycles (~3.5 s). Error bars represent the standard deviation across three repeated experiments conducted on three days in the same subject. (**c**) Statistics results (n = 10) of maximum and minimum blood pressure over time. Paired t-tests were performed between each post-exercise time point and the baseline. (**d**) Statistics results (n = 10) of maximum and minimum arterial area over time.

## 3.3 Dynamic changes in arterial active mechanics after exercise

Figure 4a shows the pressure–radius (P–r) curves from 10 pre- and post-exercise measurements in a volunteer (27 years old, male). The hysteresis loops observed in the P–r curves reflect the viscoelastic behavior of the artery, consistent with previous studies (Ghigo et al. 2017; Valdez-Jasso et al. 2011; van der Bruggen et al. 2021). Over time, the P–r curve initially shifts toward the upper-left direction (e.g., #2 vs. #0), and subsequently returning toward the baseline position. As can be explained by the theoretical curves shown in Fig. 4b (red vs. blue), an increase in $K_a$ causes an upward shift of the P–r curve. Therefore, the temporal evolution of the P–r curves observed in the experiments reflects the active regulation of arterial smooth muscle cells in response to exercise. By fitting the experimental data to the theoretical model, Fig. 4c shows the fitted values of $K_a$ over time (the corresponding fitting curves are shown in Fig. 4a). The shaded area denotes the standard deviation of the inferred values, which is substantial (s.d. / mean > 100%) and reflects a high uncertainty in the estimation of $K_a$.

The high uncertainty of $K_a$ can be theoretically explained as follows. According to Eq. (5), when blood pressure is applied on the arterial wall, the resulting equilibrium wall stress arises from both the passive stretching of fibers and the active contraction of vascular smooth muscle. As shown in Fig. 4b, two nearly identical pressure-radius relationships (dashed curve vs. blue solid curve) can be achieved by appropriately adjusting the combination of passive and active components. Therefore, it is fundamentally difficult to decouple the passive and active contributions using only the experimentally measured pressure–area data. This limitation introduces notable uncertainty in the inverse estimation of the active parameter $K_a$. To improve the reliability of the estimated parameter, we further analyzed two pressure–radius curves with identical passive components but differing levels of smooth muscle activation. As shown by the blue and red curves in Fig. 4b, at a given radius, the pressure difference between the two curves is solely attributed to the difference in active stress, which can be expressed as:

$$\Delta P = \frac{h}{r}\lambda_\theta \left[1 - \left(\frac{\lambda_m - \lambda_\theta}{\lambda_m - \lambda_0}\right)^2\right]\Delta K_a \qquad (7)$$

Eq. (7) indicates that the change in active parameter between the two states ($\Delta K_a$) is proportional to the corresponding difference in blood pressure ($\Delta P$). The estimation uncertainty of $\Delta K_a$ is influenced by the measurement errors in blood pressure, as well as by the estimation uncertainties of the smooth muscle cell-related microstructural parameters $\lambda_m$ and $\lambda_0$. Given the low inter-subject variability of $\lambda_m$ and $\lambda_0$, the parameter $\Delta K_a$ is theoretically expected to exhibit reduced uncertainty, making it more reliably inferred through model inversion than $K_a$. Based on the above theoretical analysis, the parameter $\Delta K_a$, representing the change in the active parameter at each post-exercise time point compared to baseline, was further inferred. Figure 4d presents the fitted values of $\Delta K_a$ for the same volunteer, which shows remarkably improved inversion reliability (s.d. / mean < 10%).

Table 2 summarizes the inferred active and passive mechanical parameters for all the 10 volunteers. Figure 4e presents the statistical results of $\Delta K_a$ over post-exercise time. By definition, $\Delta K_a$ is zero at baseline (time = -5 min) and therefore not displayed. A one-sample *t*-test was performed at each group to evaluate whether the post-exercise values of the active parameter $K_a$ significantly differed from the baseline value. The results indicate that arterial active mechanics remained significantly elevated compared to baseline within the first 15 minutes after exercise (*p* < 0.001 for time points #1 - #4, and *p* < 0.05 for #5). Beyond 15 minutes, no significant difference was observed, suggesting a recovery of arterial activity to baseline levels.

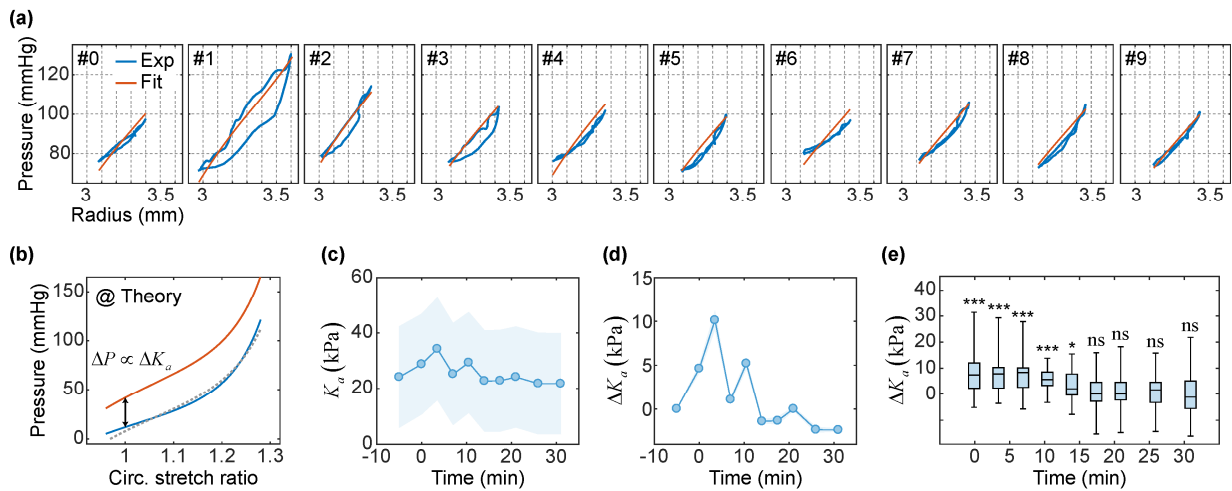

**Figure 4.** Arterial pressure–radius analysis and characterization of arterial active mechanics. (**a**) Arterial pressure–radius curves at the 10 measurement points (#0, baseline; #1 - #9, post-exercise) and their corresponding fitting curves for the representative volunteer. The radius is calculated using a circular approximation, i.e., $r = \sqrt{A/\pi}$,

where $A$ is the experimentally measured arterial area. (**b**) Theoretical relationship between pressure and circumferential stretch ratio ($\lambda_\theta = r/R$), which is equivalent to the pressure-radius relation. Blue curve: $K_a = 10$ kPa, $\mu = 5$ kPa, $\mu_t = 10$ kPa, $k_2 = 3$, $\lambda_m = 1.6$, $\lambda_0 = 0.8$, $R = 3$ mm, $H = 1$ mm, $\lambda_z = 1.1$. Red curve: $K_a = 40$ kPa, other parameters are identical to those used for the blue curve. Dashed curve: $K_a = 0$, $\mu = 20$ kPa, $k_2 = 2.5$, other parameters are identical to those used for the blue curve. (**c**) Fitting values of the active parameter $K_a$ over time for the representative volunteer. The shaded region represents the uncertainty associated with the parameter inversion. (**d**) Fitting results of change in active parameter $\Delta K_a$ over time for the same volunteer. (**e**) Statistical results (n = 10) of the change in active parameter ($\Delta K_a$) over time. A one-sample t-test was performed at each post-exercise time point.

**Table 2.** Statistics results (n = 10) of arterial passive and active mechanical parameters

| $\mu$ (kPa) | $\mu_\theta$ (kPa) | $k_2$ | $\lambda_m$ | $\lambda_0$ | $\lambda_z$ | $R$ (mm) | | |
|---|---|---|---|---|---|---|---|---|
| 4.3 ± 4.2 | 10.0 ± 3.7 | 2.4 ± 1.3 | 1.7 ± 0.1 | 0.85 ± 0.05 | 1.1 ± 0.05 | 2.8 ± 0.1 | | |
| $\Delta K_a^{(1)}$ (kPa) | $\Delta K_a^{(2)}$ | $\Delta K_a^{(3)}$ | $\Delta K_a^{(4)}$ | $\Delta K_a^{(5)}$ | $\Delta K_a^{(6)}$ | $\Delta K_a^{(7)}$ | $\Delta K_a^{(8)}$ | $\Delta K_a^{(9)}$ |
| 8.2 ± 9.6 | 8.2 ± 8.2 | 7.8 ± 7.5 | 7.0 ± 6.7 | 3.5 ± 6.3 | 1.6 ± 8.9 | 1.4 ± 7.8 | 0.9 ± 7.2 | -0.7 ± 8.0 |

# 4 Discussion

This study quantifies the temporal evolution of arterial active mechanics before and after exercise through mechanical model–based analysis of pressure–area curves. Although a relatively simple active constitutive model was adopted here for simplicity, our framework is applicable to more comprehensive active models of artery that better characterize vascular smooth muscle mechanics. To demonstrate this, we invoked two additional constitutive models, one incorporating smooth muscle fibers oriented along both longitudinal and circumferential directions (Baek et al. 2007), and a calcium-driven mechanochemical model describing actin-myosin filament interactions (Murtada et al. 2010a). In both cases, the estimated active mechanical parameters exhibited temporal trends consistent with Fig. 4d, i.e., a significant elevation persisting for approximately 15 minutes post-exercise relative to baseline (see details in Supplementary Note 3). These results indicate that the temporal variation in arterial active mechanics characterized with our method is model-independent and intrinsically linked to the active behavior of smooth muscles.

A notable observation in our study is that blood pressure normalized within about 5 minutes post-exercise, whereas arterial active mechanics required approximately 15 minutes to return to baseline. Previous studies report analogous patterns, such as delayed recovery of arterial compliance relative to blood pressure following resistance exercise (DeVan et al. 2005). The blood pressure recovery reflects rapid central autonomic adjustment (Halliwill 2001). In comparison, the delayed normalization of the active mechanical parameter suggests that smooth muscle contractility recovers more slowly. This may reflect that the intracellular pathways for active stress generation — such as $Ca^{2+}$ handling and myosin light chain dephosphorylation (Aksoy et al. 1982) — resolve more slowly than the mechanisms governing systemic pressure recovery. Further investigation is warranted to elucidate the underlying physiological mechanisms.

The CNN architecture used for image segmentation in this work has been previously reported (Chen et al. 2017; Ouyang et al. 2020). Our primary contribution lies in the use of a self-acquired and carefully annotated dataset with task-specific training tailored for arterial lumen segmentation. This trained model holds broad potential for cardiovascular health monitoring. For instance, some studies have explored continuous blood pressure monitoring using ultrasound-derived arterial diameter measurements (Wang et al. 2018a; Zhou et al. 2024). Nevertheless, such methods typically require high spatial localization accuracy and could be unsuitable for arteries with non-circular cross-sections. By applying real-time cross-sectional ultrasound imaging in combination with the CNN segmentation, more accurate and robust blood pressure monitoring could be achieved. Moreover, by implementing multiplane scanning along the arterial pathway, combined with the present image segmentation model, comprehensive 4D reconstruction of arterial geometry across space and time could become feasible. This capability may find applications in arterial biomechanics and clinical detection of localized pathologies such as stenosis.

This study incorporates several simplifications in arterial mechanical modeling. First, the membrane assumption was adopted, which is unable to account for bending and shear effects in the arterial wall. Through numerical calculations and finite element simulations, we found that for carotid arteries, where the thickness-to-radius ratio is typically around 1/3, the error introduced by using the membrane model to predict P–r curves is less than 5%. Therefore, this model approximation is considered reasonable (Supplementary Note 4 and Figure S3). Secondly, external pressure was assumed negligible, though in vivo arteries experience perivascular tissue constraints. Simulations indicate that when surrounding tissue modulus reaches 20% of the arterial wall modulus, P–r curve deviations relative to unconstrained conditions remain ≤10%, supporting this simplification (Supplementary Note 5). Thirdly, residual stress, a key determinant of vascular homeostasis (Fung 1991), has not been considered here. We further performed finite element analysis to demonstrate that, while residual stress critically modulates intramural stress distribution, its effect on the global P–r relationship is negligible (Supplementary Note 6). This is mainly because residual stress affects the bending stiffness of the structure rather than its tensile stiffness. Finally, a simplified passive constitutive model with vertically aligned fibers was used. Although arterial collagen and elastin networks exhibit complex architectures (Gasser et al. 2006; Holzapfel et al. 2015; Holzapfel and Ogden 2025), it has been shown that circumferential mechanical properties dominate P–r behavior (Jiang et al. 2022), confirming such a simplification is reasonable.

The current study has several limitations. First, the arterial wall was modeled as a single-layer structure, whereas arteries consist of three distinct layers: the intima, media, and adventitia (Holzapfel et al. 2000). The inferred mechanical properties in this study can be interpreted as an average value across layers. Utilizing higher-resolution imaging techniques may facilitate the characterization of the active and passive mechanical properties of individual arterial layers. Second, the artery was modeled as a purely elastic material, whereas arterial tissues exhibit viscoelastic behavior (Holzapfel et al. 2002). Incorporating viscoelastic constitutive models and experimental P–r loops to characterize arterial viscoelasticity requires further investigation. Third, since blood pressure and ultrasound imaging were measured separately, the phase relationship between pressure and radius waveform could not be directly determined from the experiment. This may introduce uncertainty into the P–r curve. Recent ultrasound technique enables simultaneous acquisition of pressure and radius in vivo (Jiang et al. 2025; Li et al. 2022), offering potential for accurate phase assessment between pressure and radius. Fourth, this work reliably quantifies relative changes in the active parameter (vs. baseline) but cannot accurately estimate their absolute values.

Because active smooth muscle effects and passive fiber stretch coexist in vivo, decoupling them from limited in vivo data is challenging. Characterizing the absolute value of arterial active parameter under arbitrary physiological states remains a key challenge for future research.

## 5 Conclusion

In this study, we propose an in vivo method to quantify the active mechanical parameters of arteries. The following contributions have been made: first, we trained a neural network–based segmentation model capable of achieving frame-by-frame segmentation of the arterial cross-sectional lumen. Second, we developed a biomechanical model–driven inverse approach to estimate the active mechanical parameter with low uncertainty. In vivo experiments were conducted on ten healthy volunteers to validate the method. The results revealed a significant increase in arterial active mechanics after exercise, which returned to baseline in approximately 15 minutes, substantially longer than the recovery time of blood pressure (~5 minutes). Our method and findings may provide new insights into vascular biomechanics and smooth muscle regulation, and could potentially aid in the clinical diagnosis of cardiovascular diseases.

## DATA AVAILABILITY

Data underlying the results presented in this paper are not publicly available at this time but can be obtained from the authors upon reasonable request.

## CODE AVAILABILITY

The code for the vessel segmentation neural network is available at https://github.com/JarretJiang/Vessel-Segmentation.git.

## SUPPLEMENTARY MATERIAL

All the supplementary materials are provided in a document.

## ACKNOWLEDGMENTS

We thank Ruihong Cheng for his assistance in conducting the in vivo experiments.

## GRANTS

This work was supported by the National Natural Science Foundation of China (Grants Nos. 11972206 and 12472163)

## DISCLOSURES

No conflicts of interest, financial or otherwise, are declared by the authors.

## AUTHOR CONTRIBUTIONS

**Conceptualization**: Y. J., and Y. C. **Methodology**: Y. J. **Investigation**: Y. J. **Visualization**: Y. J. **Funding acquisition:** Y. C. **Project administration**: Y. C. **Supervision**: Y. C. **Writing—original draft**: Y. J., and Y. C. **Writing—review & editing**: Y. J., and Y. C.

# REFERENCES

Aksoy M, Murphy R, Kamm K (1982) Role of Ca2+ and myosin light chain phosphorylation in regulation of smooth muscle. Am J Physiol Cell Physiol 242:C109-C116

Alpenglow JK et al. (2024) α-Adrenergic regulation of skeletal muscle blood flow during exercise in patients with heart failure with preserved ejection fraction. J Physiol 602:3401-3422

Azzopardi C, Camilleri KP, Hicks YA (2020) Bimodal automated carotid ultrasound segmentation using geometrically constrained deep neural networks. IEEE J Biomed Health Inform 24:1004-1015

Baek S, Gleason RL, Rajagopal K, Humphrey J (2007) Theory of small on large: potential utility in computations of fluid–solid interactions in arteries. Comput Methods Appl Mech Eng 196:3070-3078

Boutouyrie P, Chowienczyk P, Humphrey JD, Mitchell GF (2021) Arterial stiffness and cardiovascular risk in hypertension. Circ Res 128:864-886

Cecelja M, Chowienczyk P (2012) Role of arterial stiffness in cardiovascular disease. JRSM Cardiovasc Dis 1:1-10

Chen L-C, Papandreou G, Schroff F, Adam H (2017) Rethinking atrous convolution for semantic image segmentation. arXiv preprint arXiv:170605587

Chirinos JA (2012) Arterial stiffness: basic concepts and measurement techniques. J Cardiovasc Transl Res 5:243-255

Chirinos JA, Segers P, Hughes T, Townsend R (2019) Large-artery stiffness in health and disease: JACC state-of-the-art review. J Am Coll Cardiol 74:1237-1263

Couade M et al. (2010) Quantitative assessment of arterial wall biomechanical properties using shear wave imaging. Ultrasound Med Biol 36:1662-1676

D Lefrandt J, J Smit A, J Zeebregts C, OB Gans R, H Hoogenberg K (2010) Autonomic dysfunction in diabetes: a consequence of cardiovascular damage. Curr Diabetes Rev 6:348-358

DeVan AE, Anton MM, Cook JN, Neidre DB, Cortez-Cooper MY, Tanaka H (2005) Acute effects of resistance exercise on arterial compliance. J Appl Physiol 98:2287-2291

Franchini G, Breslavsky ID, Giovanniello F, Kassab A, Holzapfel GA, Amabili M (2022) Role of smooth muscle activation in the static and dynamic mechanical characterization of human aortas. Proc Natl Acad Sci USA 119:e2117232119

Fung YC (1991) What are the residual stresses doing in our blood vessels? Ann Biomed Eng 19:237-249

Gade J-L, Thore C-J, Sonesson B, Stålhand J (2021) In vivo parameter identification in arteries considering multiple levels of smooth muscle activity. Biomech Model Mechanobiol 20:1547-1559

Gasser TC, Ogden RW, Holzapfel GA (2006) Hyperelastic modelling of arterial layers with distributed collagen fibre orientations. J R Soc Interface 3:15-35

Ghigo AR, Wang X-F, Armentano R, Fullana J-M, Lagrée P-Y (2017) Linear and nonlinear viscoelastic arterial wall models: application on animals. J Biomech Eng 139:011003

Golemati S, Stoitsis J, Sifakis EG, Balkizas T, Nikita KS (2007) Using the Hough transform to segment ultrasound images of longitudinal and transverse sections of the carotid artery. Ultrasound Med Biol 33:1918-1932

Guerrero J, Salcudean SE, McEwen JA, Masri BA, Nicolaou S (2007) Real-time vessel segmentation and tracking for ultrasound imaging applications. IEEE Trans Med Imaging 26:1079-1090

Halliwill JR (2001) Mechanisms and clinical implications of post-exercise hypotension in humans. Exerc Sport Sci Rev 29:65-70

Harvey A, Montezano AC, Touyz RM (2015) Vascular biology of ageing—Implications in hypertension. J Mol Cell Cardiol 83:112

Holzapfel GA, Gasser TC, Ogden RW (2000) A new constitutive framework for arterial wall mechanics and a comparative study of material models. J Elast 61:1-48

Holzapfel GA, Gasser TC, Stadler M (2002) A structural model for the viscoelastic behavior of arterial walls: continuum formulation and finite element analysis. Eur J Mech A-Solids 21:441-463

Holzapfel GA, Niestrawska JA, Ogden RW, Reinisch AJ, Schriefl AJ (2015) Modelling non-symmetric collagen fibre dispersion in arterial walls. J R Soc Interface 12:20150188

Holzapfel GA, Ogden RW (2010) Constitutive modelling of arteries. Proc R Soc A-Math Phys Eng Sci 466:1551-1597

Holzapfel GA, Ogden RW (2025) Modeling the biomechanical properties of soft biological tissues: Constitutive theories. Eur J Mech A-Solids 112:105634

Humphrey J, Na S (2002) Elastodynamics and arterial wall stress. Ann Biomed Eng 30:509-523

Jiang Y et al. (2025) Simultaneous imaging of bidirectional guided waves probes arterial mechanical anisotropy, blood pressure, and stress synchronously. Sci Adv 11:eadv5660

Jiang Y, Zheng Y, Li G-Y, Zhang Z, Yin Z, Xu W, Cao Y (2022) Probing the Mechanical Properties of Large Arteries by Measuring Their Deformation In Vivo with Ultrasound. Ultrasound Med Biol 48:1033-1044

Jurik R, Żebrowska A, Stastny P (2021) Effect of an acute resistance training bout and long-term resistance training program on arterial stiffness: a systematic review and meta-analysis. J Clin Med 10:3492

Kelly R, Fitchett D (1992) Noninvasive determination of aortic input impedance and external left ventricular power output: a validation and repeatability study of a new technique. J Am Coll Cardiol 20:952-963

Khamdaeng T, Luo J, Vappou J, Terdtoon P, Konofagou E (2012) Arterial stiffness identification of the human carotid artery using the stress–strain relationship in vivo. Ultrasonics 52:402-411

Laurent S et al. (2006) Expert consensus document on arterial stiffness: methodological issues and clinical applications. Eur Heart J 27:2588-2605

Li G-Y, Jiang Y, Zheng Y, Xu W, Zhang Z, Cao Y (2022) Arterial stiffness probed by dynamic ultrasound elastography characterizes waveform of blood pressure. IEEE Trans Med Imaging 41:1510-1519

Masson I, Beaussier H, Boutouyrie P, Laurent S, Humphrey JD, Zidi M (2011) Carotid artery mechanical properties and stresses quantified using in vivo data from normotensive and hypertensive humans. Biomech Model Mechanobiol 10:867-882

Masson I, Boutouyrie P, Laurent S, Humphrey JD, Zidi M (2008) Characterization of arterial wall mechanical behavior and stresses from human clinical data. J Biomech 41:2618-2627

Mitchell GF et al. (2010) Arterial stiffness and cardiovascular events: the Framingham Heart Study. Circulation 121:505-511

Murtada S-I, Kroon M, Holzapfel GA (2010a) A calcium-driven mechanochemical model for prediction of force generation in smooth muscle. Biomech Model Mechanobiol 9:749-762

Murtada S-I, Kroon M, Holzapfel GA (2010b) Modeling the dispersion effects of contractile fibers in smooth muscles. J Mech Phys Solids 58:2065-2082

Noble JA, Boukerroui D (2006) Ultrasound image segmentation: a survey. IEEE Trans Med Imaging 25:987-1010

Ogden R (2001) Elements of the theory of finite elasticity. In: Fu Y, Ogden R (eds) Nonlinear elasticity: theory and applications. Cambridge University Press, Cambridge, pp 1-57

Ouyang D et al. (2020) Video-based AI for beat-to-beat assessment of cardiac function. Nature 580:252-256

Pewowaruk RJ, Gepner AD (2022) Smooth muscle tone alters arterial stiffness: the importance of the extracellular matrix to vascular smooth muscle stiffness ratio. J Hypertens 40:512-519

Rachev A, Hayashi K (1999) Theoretical study of the effects of vascular smooth muscle contraction on strain and stress distributions in arteries. Ann Biomed Eng 27:459-468

Reesink KD, Spronck B (2019) Constitutive interpretation of arterial stiffness in clinical studies: a methodological review. Am J Physiol-Heart Circul Physiol 316:H693-H709

Rizzoni D, Rosei EA (2006) Small artery remodeling in hypertension and diabetes. Curr Hypertens Rep 8:90-95

Salvi P et al. (2019) Noninvasive estimation of aortic stiffness through different approaches: comparison with intra-aortic recordings. Hypertension 74:117-129

Schmitz A, Böl M (2011) On a phenomenological model for active smooth muscle contraction. J Biomech 44:2090-2095

Schofield I, Malik R, Izzard A, Austin C, Heagerty A (2002) Vascular structural and functional changes in type 2 diabetes mellitus: evidence for the roles of abnormal myogenic responsiveness and dyslipidemia. Circulation 106:3037-3043

Segers P, Rietzschel ER, Chirinos JA (2020) How to measure arterial stiffness in humans. Arterioscler Thromb Vasc Biol 40:1034-1043

Spronck B et al. (2015) A constitutive modeling interpretation of the relationship among carotid artery stiffness, blood pressure, and age in hypertensive subjects. Am J Physiol-Heart Circul Physiol 308:H568-H582

Stålhand J, Klarbring A, Holzapfel GA (2008) Smooth muscle contraction: mechanochemical formulation for homogeneous finite strains. Prog Biophys Mol Biol 96:465-481

Studinger P, Lénárd Z, Kováts Z, Kocsis L, Kollai M (2003) Static and dynamic changes in carotid artery diameter in humans during and after strenuous exercise. J Physiol 550:575-583

Teixeira R, Vieira MJ, Gonçalves A, Cardim N, Gonçalves L (2016) Ultrasonographic vascular mechanics to assess arterial stiffness: a review. Eur Heart J Cardiovasc Imaging 17:233-246

Touyz RM, Alves-Lopes R, Rios FJ, Camargo LL, Anagnostopoulou A, Arner A, Montezano AC (2018) Vascular smooth muscle contraction in hypertension. Cardiovasc Res 114:529-539

Valdez-Jasso D, Bia D, Zócalo Y, Armentano RL, Haider MA, Olufsen MS (2011) Linear and nonlinear

viscoelastic modeling of aorta and carotid pressure–area dynamics under in vivo and ex vivo conditions. Ann Biomed Eng 39:1438-1456

Van Bortel LM et al. (2012) Expert consensus document on the measurement of aortic stiffness in daily practice using carotid-femoral pulse wave velocity. J Hypertens 30:445-448

van der Bruggen MM et al. (2021) An integrated set-up for ex vivo characterisation of biaxial murine artery biomechanics under pulsatile conditions. Sci Rep 11:2671

Wagner H, Humphrey J (2011) Differential passive and active biaxial mechanical behaviors of muscular and elastic arteries: basilar versus common carotid. J Biomech Eng 133:051009

Wang C et al. (2018a) Monitoring of the central blood pressure waveform via a conformal ultrasonic device. Nat Biomed Eng 2:687-695

Wang D, Tan D, Liu L (2018b) Particle swarm optimization algorithm: an overview. Soft Comput 22:387-408

Wang DC, Klatzky R, Wu B, Weller G, Sampson AR, Stetten GD (2009) Fully automated common carotid artery and internal jugular vein identification and tracking using B-mode ultrasound. IEEE Trans Biomed Eng 56:1691-1699

Zakrzewski AM, Anthony BW (2017) Noninvasive blood pressure estimation using ultrasound and simple finite element models. IEEE Trans Biomed Eng 65:2011-2022

Zhang Y, Zhang Y-J, Ye W, Korivi M (2021) Low-to-moderate-intensity resistance exercise effectively improves arterial stiffness in adults: evidence from systematic review, meta-analysis, and meta-regression analysis. Front Cardiovasc Med 8:738489

Zhou S et al. (2024) Clinical validation of a wearable ultrasound sensor of blood pressure. Nat Biomed Eng:1-17

Zulliger MA, Rachev A, Stergiopulos N (2004) A constitutive formulation of arterial mechanics including vascular smooth muscle tone. Am J Physiol-Heart Circul Physiol 287:H1335-H1343

# Supplementary materials

# In Vivo Quantification of Arterial Active Mechanics Using Deep Learning-Assisted Pressure-Area Analysis

Yuxuan Jiang, Yanping Cao[*]

*Institute of Biomechanics and Medical Engineering, AML, Department of Engineering Mechanics,*

*Tsinghua University, Beijing 100084, PR China*

[*]Corresponding author: Yanping Cao
Email address: caoyanping@tsinghua.edu.cn; Tel: 86-10-62772520; Fax: 86-10-62781284.

**Supplementary Figures**

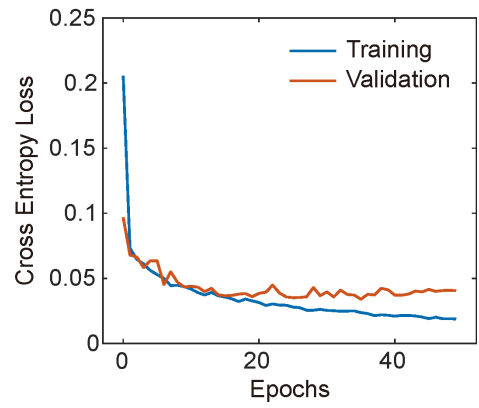

**Figure S1.** Cross entropy loss for vessel segmentation during training on the training and validation datasets.

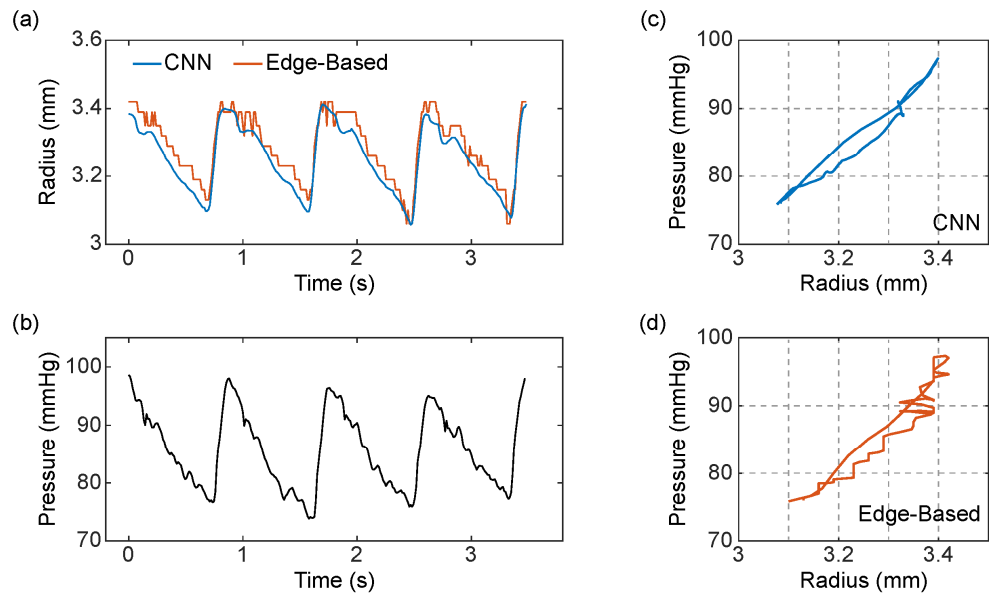

**Figure S2.** Comparison of pressure-radius curves derived from the CNN-based and edge-based segmentation methods. CNN, Convolutional neural network. **(a)** Arterial radius over time, derived from the two segmentation methods. The radius was calculated from the segmented area by assuming a circular shape, i.e. $r = \sqrt{A/\pi}$. **(b)** Blood pressure over time. **(c)** Pressure-radius (P-r) curve derived from the pressure data and CNN-based radius data. **(d)** P-r curve derived from the pressure data and Edge-based radius data. Due to the significant noise in the radius data obtained from the edge-based method, the resulting P–r curve exhibits considerable measurement errors compared to the CNN-based approach.

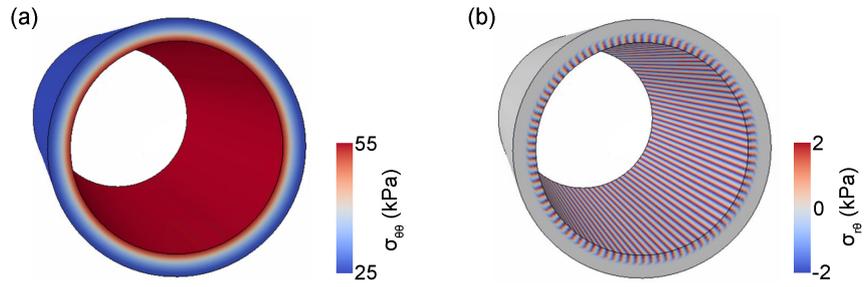

**Figure S3.** Finite element analysis of stress components in the arterial wall. **(a)** Distribution of the circumferential normal stress $\sigma_{\theta\theta}$. **(b)** Distribution of the shear stress in the radial-circumferential plane $\sigma_{r\theta}$. Model parameters: inner radius in the initial state $R_i$ = 2.5 mm; outer radius in the initial state $R_o$ = 3.5 mm; pressure $P$ = 8 kPa; constitutive parameters $\mu$ = 20 kPa, $\mu_z = \mu_\theta$ = 40 kPa, $k_2$ = 0.7.

## Supplementary Note 1. Comparison of neural network training using 2 and 4 annotated frames

The neural network is trained in a weakly supervised manner in this work, where only a few manually annotated frames are selected from each video (~300 frames). Here we investigated how the number of annotated frames affects network performance. Figure S4a shows the strategy used in the main text, where two frames per video—end-diastole and peak-systole—were annotated for training. Figure S4b illustrates an alternative approach where four frames (end-diastole, peak-systole, and two additional cardiac phases) were annotated for each video. Apart from the difference in the number of annotated frames used for supervision, the two training schemes shared the same datasets. As shown in Fig. S4c, the segmentation results from the 2-frame and 4-frame supervision strategies are similar, in terms of both per-frame segmentation region and the temporal evolution of lumen area. The Dice coefficients on the test dataset were also close (0.959 for 2-frame vs. 0.957 for 4-frame supervision). These findings suggest that increasing the number of annotated frames does not lead to a significant improvement in segmentation performance. This is primarily because the current neural network lacks temporal modeling. As a result, training is more influenced by static B-mode image features specific to each subject, while changes in lumen area over time does not substantially alter the image pattern and therefore contributes less to prediction performance.

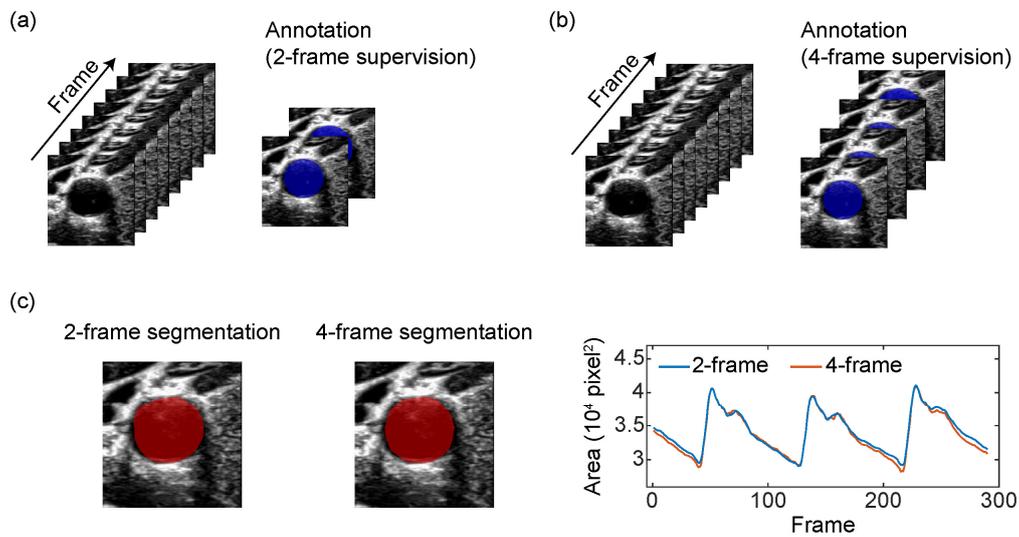

**Figure S4.** Neural network training using 2 and 4 annotated frames. **(a)** Each video contains two annotated frames. **(b)** Each video contains four annotated frames. **(c)** Left: The segmentation result under the 2-frame supervision strategy. One frame from a test dataset video was selected for demonstration. Middle: The segmentation result under the 4-frame supervision strategy. Right: Temporal (frame-by-frame) segmented areas from the two strategies.

**Supplementary Note 2. An edge-based method for arterial lumen segmentation**

Figure S5 illustrates an edge-based method for arterial segmentation. First, edge enhancement is applied to the B-mode image of the arterial cross-section using the Canny edge detector. Then, circular ring templates of varying diameters (with a fixed wall thickness of 10 pixels) are convolved with the edge-enhanced image. The ring yielding the highest convolution response is selected as the segmented vessel boundary. The area of the circle is then taken as the segmented lumen area.

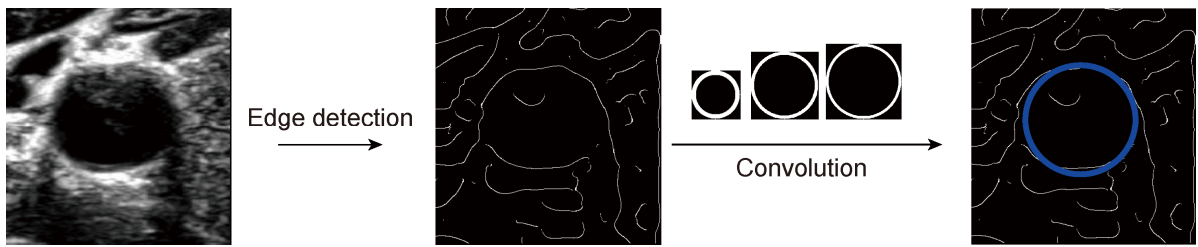

**Figure S5.** A conventional edge-based method for arterial lumen segmentation.

# Supplementary Note 3. Characterizing arterial active mechanics using various active constitutive models

Here, we incorporate more advanced active constitutive models into the P–r curve inversion framework and investigate how the corresponding active parameters vary before and after exercise. Two active constitutive models are considered below, both of which provide a more accurate representation of smooth muscle cell compared to the model used in the main text. The first active model (marked as 'Model A') accounts for smooth muscle distribution in both circumferential and longitudinal directions. The strain energy function remains as defined in Eq. (4), i.e.

$$W^{(a,\text{Model-A})} = K_a \left[ \lambda_f + \frac{1}{3} \frac{(\lambda_m - \lambda_f)^3}{(\lambda_m - \lambda_0)^2} \right] \tag{4}$$

where the fiber stretch ratio ($\lambda_f$) is defined by

$$\lambda_f = \sqrt{\frac{1}{2}(\lambda_\theta^2 + \lambda_z^2)} \tag{S1}$$

The second active model (marked as 'Model B') incorporates the microscale physiology of smooth muscle contraction — the interaction of sliding actin and myosin filaments (Murtada et al. 2010). This mechanochemical model links the level of active contraction to intracellular calcium concentration, providing a more physiologically meaningful description of smooth muscle behavior. The strain energy function is formulated as

$$W^{(a,\text{Model-B})} = \frac{\mu_a}{2}(n_c + n_D)(\lambda_\theta + \bar{u}_{rs} - 1)^2 \tag{S2}$$

where $\mu_a$ denotes an active parameter with the same dimension as a modulus. $n_c$ and $n_c$ are the concentrations related to the two attached states. $\bar{u}_{rs}$ (~ 0.05) denotes the normalized relative sliding between the thick and thin filaments. During inversion, we treated $\mu_a(n_c + n_D)$ as a single parameter, representing the contractile strength of the smooth muscle cells, similar in the meaning to $K_a$.

Figure S6a shows the inferred active parameter using Model A. The differential value $\Delta K_a$, rather than the absolute value $K_a$, is presented due to its better inversion stability. Figure S6b shows the inferred active parameter $\Delta(\mu_a n_c + \mu_a n_D)$ using Model B. Similar to the trend shown in Fig. 4d, the inferred active parameters remain significantly elevated compared to baseline within approximately 15 minutes after exercise, and subsequently shows no statistical difference from baseline, indicating a recovery of arterial activity to resting levels. The values of the parameters in Fig. S6 and Fig. 4d differ primarily due to the different mathematical forms of the strain energy functions used. In summary, the proposed inversion framework is applicable to a broad range of active models. The post-exercise trend of arterial activation revealed in this study is independent of the specific choice of constitutive model, suggesting a generalizable physiological pattern.

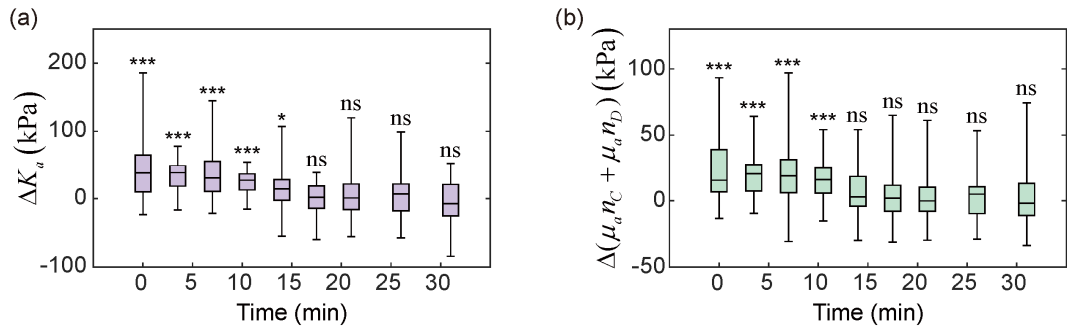

**Figure S6.** Statistical results (n = 10) of the change in active parameters over time using different active constitutive models. **(a)** Temporal evolution of active parameter $\Delta K_a$ using Model A. **(b)** Temporal evolution of active parameter $\Delta(\mu_a n_C + \mu_a n_D)$ using Model B.

## Supplementary Note 4. Comparison of thick-walled model and thin-walled model

In this study, we adopted the thin-walled model to predict the relationship between arterial blood pressure and area. The thin-walled assumption could introduce some estimation errors, as the thick-to-radius ratio for the carotid artery is approximately 0.3 according to our experiments. Therefore, we compare the thin-walled model to the thick-walled model to quantitatively study this estimation error. Consider an incompressible thick-walled tube subjected to intraluminal inflation under pressure $P$. The inner and outer radius of the tube in the unloaded state are denoted as $R_i$ and $R_o$, respectively. The thickness is $H$ ($= R_i - R_0$). The corresponding radii and thickness at the loaded state are marked as $r_i$, $r_o$, and $h$, respectively. The relationship between pressure and arterial radius is (Ogden 2001)

$$P = \int_{\lambda_{\theta,o}}^{\lambda_{\theta,i}} (\lambda^2 \lambda_z - 1)^{-1} \frac{\partial W}{\partial \lambda} d\lambda \tag{S3}$$

where $\lambda_\theta$ and $\lambda_z$ denotes the circumferential and longitudinal stretch ratio, respectively. $\lambda_{\theta,o} (= r_o/R_o)$ and $\lambda_{\theta,i} (= r_i/R_i)$ denote the circumferential stretch ratio at the outer radius and the inner radius, respectively. $W$ is the strain energy function of the material. With the incompressibility condition, they are related in the form of

$$\lambda_{\theta,o}^2 = \frac{R_i^2}{R_o^2} \lambda_{\theta,i}^2 + \left(1 - \frac{R_i^2}{R_o^2}\right) \frac{1}{\lambda_z} \tag{S4}$$

Figure S7 compares the pressure-radius curves predicted by the thin-walled and thick-walled models. The range of pressure is chosen within the physiological range, and the thickness-to-radius ratio is set as 1/3. All the material and geometric parameters used in Eq. (S3) are chosen based on the in vivo experiments. As shown in Fig. S7a, under the same pressure, the radius predicted by the thin-walled model is larger, indicating a lower structural stiffness of tube. This is reasonable, as the thin-walled model neglects the bending stiffness of the tube. With the increase of pressure, the difference between the two curves grows larger. As shown in Fig. S7b, the relative difference remains below 2.5% when the blood pressure is below 180 mmHg. This indicates that for the carotid artery in our study, using the thin-walled model to predict the pressure-radius relationship is reasonable, with an estimation error of less than 3%.

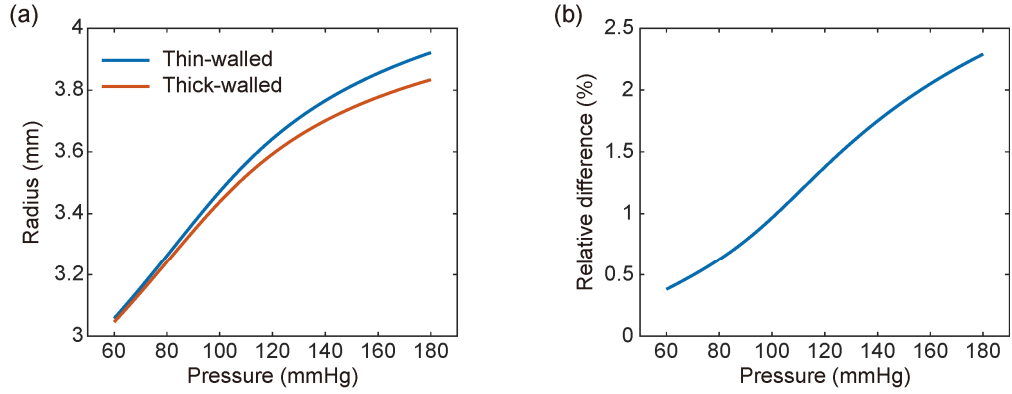

**Figure S7**. Comparison of the thin-walled and thick-walled model. **(a)** Relationship of pressure and radius predicted by the two models. Parameters: $\mu = 10$ kPa, $\mu_\theta = 10$ kPa, $k_2 = 2$, $K_a = 50$ kPa, $\lambda_m = 1.6$, $\lambda_0 = 0.8$, $\lambda_z = 1.1$. Thick-walled model: $R_i = 2.5$ mm, $R_o = 3.5$ mm, $H = 1$ mm. Thin-walled model: $R = 3$ mm, $H = 1$ mm. The radius plotted on the y-axis refers to the central radius, $(R_i+R_o)/2$, for the thick-walled model. **(b)** Relative difference of the two curves. Relative difference = $|r_2 - r_1| / r_1$, where $r_2$ and $r_1$ denote the radius predicted by the thick-walled and thin-walled model, respectively.

**Supplementary Note 5. Influence of surrounding tissues on arterial deformation**

In vivo, arteries are not only subjected to internal blood pressure but also constrained by surrounding tissues. Figure S8a illustrates a mechanical model that accounts for the surrounding tissues. The pressure–deformation relationship of the artery is described as

$$P = \frac{H}{R}\frac{1}{\lambda_\theta \lambda_z}\frac{\partial W}{\partial \lambda_\theta} + \int_{\frac{1}{\sqrt{\lambda_z}}}^{\lambda_\theta}(\lambda^2 \lambda_z - 1)^{-1}\frac{\partial W_s}{\partial \lambda}\,\mathrm{d}\lambda \tag{S5}$$

where $\lambda_\theta$ and $\lambda_z$ denote the circumferential and longitudinal stretch ratio of the artery, respectively. $W$ denotes the strain energy function of the artery (i.e. Eqs. (2) - (4) in the main text). $W_s$ denotes the strain energy function of the surrounding tissues. The surrounding tissue was modeled using the Fung – Demiray constitutive model (Demiray 1972), which has been widely used to describe the nonlinear stiffening of soft biological tissues. The strain energy function is given by:

$$W_s = \frac{\mu_s}{2b_s}\{\exp[b_s(I_1 - 3)] - 1\} \tag{S6}$$

where $\mu_s$ denotes the shear modulus of the tissue. $b_s$ is a dimensionless parameter denoting the nonlinear stiffening of the tissue. $I_1$ is the first invariant of the strain tensor. Using Eq. (S5), Fig. S8b compares the pressure-radius relationship with and without the inclusion of surrounding tissues. The presence of surrounding tissue results in a smaller arterial radius under the same pressure, which is reasonable due to its constraining effect. When the shear modulus ratio of surrounding tissues to artery ($\mu_s/\mu_\theta$) is within 20%—as expected in vivo—the relative difference in the P–r curve is less than 10%. Therefore, omitting the surrounding constraint in mechanical modeling does not introduce significant error.

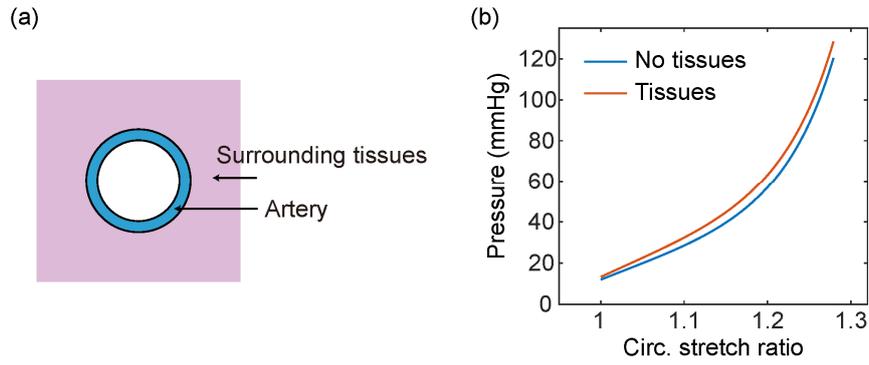

**Figure S8.** Influence of surrounding tissues on the pressure-radius relationship. **(a)** Schematic of the artery and its surrounding tissues. **(b)** Relationship of pressure and circumferential stretch ratio ($\lambda_\theta$) with and without surrounding tissues, respectively. Parameters of the artery: $\mu = 5$ kPa, $\mu_\theta = 10$ kPa, $k_2 = 3$, $K_a = 10$ kPa, $\lambda_m = 1.6$, $\lambda_0 = 0.8$, $\lambda_z = 1.1$, $H/R = 1/3$. Parameters of the surrounding tissues: $\mu_s = 2$ kPa, $b_s = 1$.

**Supplementary Note 6. Influence of residual stress on arterial deformation**

In order to account for the arterial residual stress, three configurations are defined in Fig. S9a: (i) the stress-free state, in which the artery opens into a sector after a radial cut; (ii) the unloaded state, in which the artery is closed but retains residual circumferential stress; and (iii) the loaded state, in which the artery is subjected to internal pressure and a prescribed axial stretch. Finite element analysis (Abaqus 6.14, Dassault Systemes, USA) was performed to study the effect of residual stress on the arterial deformation and stress distribution. Approximately 70,000 solid elements (C3D8R) were used to discretize the structure, and numerical convergence was verified by mesh refinement. Figure S9b shows the simulated results of stress distribution when considering the residual stress, while Fig. S9c shows the stress distribution in arterial wall without residual stress. Accounting for residual stress results in a more uniform stress distribution through the arterial wall, in agreement with physiological observations. Figure S9d compares the pressure-radius (P-r) relation calculated from the two models. Interestingly, the two curves are very close (relative difference < 1%), which indicates that simplifying the arterial residual stress would not introduce obvious error in estimating the P-r relation.

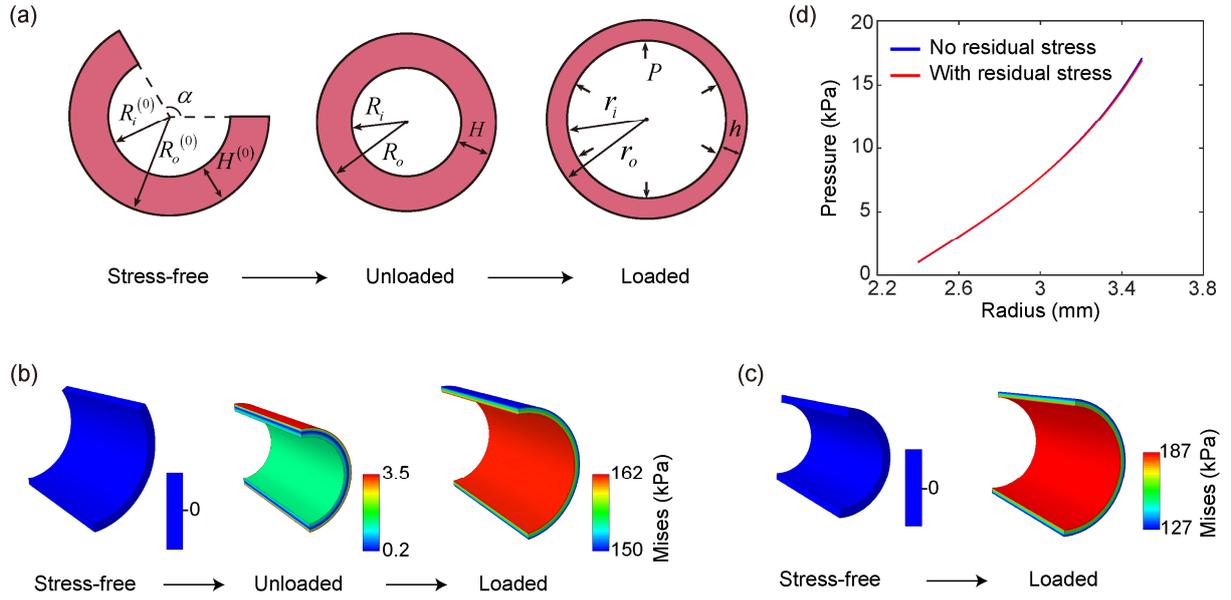

**Figure S9.** Finite element analysis of the arterial wall with and without residual stress. **(a)** Three configurations of the arterial wall with residual stress, including stress-free state, unloaded state (with residual stress and zero pressure), and loaded state (with pressure). $\alpha$ denotes the opening angle. $R_i^{(0)}$, $R_i$, and $r_i$ denote the inner radius in the corresponding states. $R_o^{(0)}$, $R_o$, and $r_o$ denote the outer radius in the corresponding states. $H^{(0)}$, $H$, and $h$ denote the wall thickness in the corresponding states. **(b)** Stress distribution of the model with residual stress. Model parameters: $R_i^{(0)}$ = 4.18 mm, $R_o^{(0)}$ = 4.72 mm, $\alpha = 2\pi/3$, $P$ = 16 kPa, $\mu$ = 10 kPa, $\mu_z = \mu_\theta = 40$ kPa, $k_2 = 1$. **(c)** Stress distribution of the model without residual stress. Model parameters: $R_i$ = 2.7 mm, $R_o$ = 3.24 mm. Other parameters are the same as those used in (b). **(d)** Comparison of pressure-radius relation predicted by the models with and without residual stress.

**Supplementary Videos**

**Video S1**. Representative example of frame-by-frame arterial segmentation results using the CNN-based method. The example presented here is consistent with the case illustrated in Fig. 2b.

**Video S2.** Frame-by-frame arterial segmentation for a low-SNR case, comparing the CNN-based method with the conventional edge-based approach. The example presented here is consistent with the case illustrated in Fig. 2d.

**Video S3.** Frame-by-frame arterial segmentation for a case with global motion, comparing the CNN-based method with the conventional edge-based approach. The example presented here is consistent with the case illustrated in Fig. 2e.

**References**

Demiray H (1972) A note on the elasticity of soft biological tissues. J Biomech 5:309-311

Murtada S-I, Kroon M, Holzapfel GA (2010) A calcium-driven mechanochemical model for prediction of force generation in smooth muscle. Biomech Model Mechanobiol 9:749-762

Ogden R (2001) Elements of the theory of finite elasticity. In: Fu Y, Ogden R (eds) Nonlinear elasticity: theory and applications. Cambridge University Press, Cambridge, pp 1-57